\documentclass{aa}
\usepackage{graphicx}
\begin{document}
\title{Cosmic rays from Galactic pulsars}

   \author{W. Bednarek \& M. Bartosik}

   \offprints{bednar@fizwe4.fic.uni.lodz.pl}

   \institute{Department of Experimental Physics, University of \L \'od\'z,
        ul. Pomorska 149/153, 90-236 \L \'od\'z, Poland
             }

   \date{Received ; accepted}

   \abstract{
We calculate energy spectra and mass composition of cosmic rays accelerated by the 
galactic population of pulsars during their radio and gamma-ray phase.   
It is assumed that a significant part of the pulsar rotational energy 
is lost on acceleration of iron nuclei extracted from the surface of the neutron star.
The nuclei are accelerated, at first when passing the outer gap of the inner pulsar 
magnetosphere and later in the pulsar wind zone, to energies corresponding to 
50\% of the total potential drop through the polar cap region of the neutron star.
We calculate energy spectra of the nuclei injected from the pulsar wind nebulae into 
the Galaxy including different
energy loss processes of nuclei during their propagation in the pulsar magnetosphere 
and the expanding nebula, their fragmentation in collisions with radiation
and matter, adiabatic energy losses, and escape from the nebula.
Several models proposed for the distribution of the initial parameters 
of the galactic pulsar population are considered. It is shown that the best description 
of the observed cosmic ray spectrum and the mass composition between a few $10^{15}$ eV 
and a few $10^{18}$ eV is obtained for the model
B of Lorimer et al., in which the logs of initial pulsar periods and surface magnetic 
fields are given by the Gaussian distributions with the average values of
$<log P[ms]>=2.6$ and $<log B[G]>=12.3$, respectively. In order to supply the cosmic rays
into the Galaxy with the required rate, the product of the efficiency of cosmic ray 
acceleration by pulsars and their birth rate should be about $10^{-2}$ yr$^{-1}$.

\keywords{supernova remnants: pulsars: general -- cosmic rays}
   }

   \maketitle
%

\section{Introduction}

The origin of cosmic rays (CRs) with energies around and above $\sim 10^{15}$ eV, 
the knee region, is still far from being understood. It is widely accepted that the 
lower energy CRs, below the knee, are accelerated at large scale shock waves created
in the interstellar medium by supernova explosions. It is argued that in some
favorite conditions the particles might be also accelerated above
$10^{15}$ eV by this same supernova shock mechanism if, e.g.
multiple supernovae exploded in the star forming regions - superbubbles
(Cesarsky \& Montmerle~1983, Bykov~2001), a supernova
explodes in its dense wind (Biermann~1993, Stanev et al.~1993), the magnetic field
at the shock front is non-linearly amplified by CRs (Lucek \& Bell~2000,
Bell \& Lucek~2001, Drury et al.~2003), or superheavy nuclei are accelerated
(H\"orandel~2003). The acceleration of particles to such energies might also 
occur in a large scale terminal shock of the galactic wind (Jokipii \& Morfill~1985).
Other models propose that the observed cosmic rays has extragalactic origin.
For example, Protheroe \& Szabo (1992) consider their acceleration in the 
shocks in the accretion flows in active galactic nuclei, and Vietri (1995) and Waxman 
(1995) argue for acceleration on relativistic shocks of gamma-ray bursts (GRBs).
This last idea has been considered recently in a more detail by Wick et al. (2004), who
propose the origin of cosmic rays from $\sim 10^{14}$ eV up to the highest energies
in the GRBs including the events which occurred in our own Galaxy with typical rate of
$(3-10)\times 10^{-6}$ per yr. For the extensive review of recent models of the 
cosmic ray acceleration we refer to H\"orandel~(2004).

Another acceleration mechanism was proposed long time ago, soon after the pulsar
discovery. It postulated acceleration of CRs above the knee region
in the pulsar wind zone in large amplitude electromagnetic waves generated by rotating
neutron star (Ostriker \& Gunn~1969,
Karaku\l a et al.~1974). The contribution of particles accelerated by pulsars
to the observed cosmic ray spectrum has been more recently
discussed in detail by Cheng \& Chi~(1996), Bednarek \& Protheroe~(2002),
Giller \& Lipski~(2002). For example, Bednarek \& Protheroe~(2002) estimate the
contribution of heavy nuclei accelerated in the pulsar outer gaps (Cheng et al.~1986)
to energies above the knee region. Giller \& Lipski~(2002) derive the
initial parameters of the pulsar population inside the Galaxy required to explain the
observed shape of the CR spectrum and its intensity.
Note that also CRs with extremely high energies, i.e. above the ankle at
$\sim 10^{18}$ eV, might be accelerated in the wind regions of
pulsars with super strong surface magnetic fields, objects called magnetars
(Blasi et al.~2000, Arons~2003).

The models postulating acceleration of CRs between the knee and the
ankle locally in our Galaxy obtained strong support by recent observation of the
cosmic ray anisotropies from the direction of the Galactic Centre and the Cygnus region
(Hayashida et al.~1999, Bellido et al.~2001). The AGASA experiment
(Hayashida et al.~1999) reported the
excesses of CRs in the narrow energy range close to $\sim 10^{18}$ eV with the
significance of $4.5\sigma$ and $3.9\sigma$ for the Galactic Centre and the Cygnus region,
respectively.

Another information on the origin of CRs can be reached by measuring its mass
composition. Most of the experiments suggest that the average mass composition of
CRs is relatively light below the knee (e.g. Burnett et al.~1990,
Apanasenko et al.~2001) and becomes heavier above the knee (e.g. Glasmacher et
al.~1999, Arqueros et al.~2000, Fowler et al.~2001, Shirasaki et al.~2001,
Ave et al.~2003). Recent preliminary results obtained from the analysis of the 
KASCADE data confirm heavier composition above the knee and show 
evidences of the rigidity dependent break
in the spectra of specific groups of nuclei (e.g. Roth et al.~2003).
The mass composition seems to decline again to lighter between $10^{17}-10^{18}$ eV,
i.e. at the region of the second knee,
according to some experiments (Bird et al.~1993, Yoshida et al.~1995, Dawson et al.~1998), 
or may be still quite heavy in this energy range as it is obtained in the analysis of the 
Haverah Park data which show bi-modal composition composed of one third of protons and 
two thirds of iron (Abu-Zayyad et al.~2001). Recent analysis of the Vulcano Range data 
also suggest heavy composition composed of 88$\%$ of the iron 
nuclei in the bi-modal proton-iron composition at a median energy of $10^{18}$ eV  
(Dova et al.~2003). The change of the cosmic ray mass composition at the knee region
is often explained as a result of rigidity dependent acceleration causing heavy nuclei
start to dominate at higher energies (Peters~1961). The heavy nuclei enrichment of the
galactic CRs might be also due to
propagation effects in the galactic disk and halo (e.g. Maurin et al.~2003).
Moreover, a contribution of another type of sources to CRs above the knee
region, e.g. pulsars, may also increase its average mass.

In this paper we consider the last possibility in more details.
We calculate the energy spectra of different types of nuclei injected by the pulsars
into the Galaxy, applying a model for their injection and propagation in the
expanding pulsar wind nebula (PWNa) surrounding a young pulsar.
The model takes into account the energy losses and escape conditions of nuclei
during the expansion of the nebula. Similar model has been recently applied for modelling
the $\gamma$-ray emission from the PWNe
(Bednarek \& Bartosik~2003). We estimate the contribution of nuclei, escaping from
the PWNe, to the observed cosmic ray spectrum above the knee for different models of the
pulsar population in the Galaxy. The results of calculations are compared
with the reports on the mass composition in this energy region.

\section{The energy spectra of nuclei injected from the pulsar wind nebulae}

In this section we calculate spectra of nuclei
accelerated by young pulsars and injected from the pulsar wind
nebulae into the Galactic medium. We define the acceleration mechanism of nuclei,
consider the effects of their propagation in the expanding nebula, and discuss
the results for different models of the galactic pulsar population.

\subsection{Acceleration of nuclei by the pulsar}

It is likely that rotating magnetospheres of neutron stars can
accelerate not only leptons but also heavy nuclei, extracted from
positively charged polar cap regions. In fact different aspects of high energy
phenomena around pulsars may need the presence of heavy nuclei.
For example, a flow of iron nuclei, coexisting with an outflowing electron-positron
plasma, is postulated in order to explain the change in the drift direction of the radio
subpulses (Gil et al.~2003).
The presence of heavy nuclei can also explain morphological features of the Crab Nebula
and the appearance of extremely energetic leptons inside the nebula,
accelerated as a result of resonant scattering of positrons and electrons by heavy
nuclei (Hoshino et al. 1992, Gallant \& Arons 1994).
From normalization to the observations of the Crab pulsar, Arons and collaborators
(e.g. see Arons~1998) postulate that the Lorentz factors of iron nuclei, accelerated
somewhere in the inner magnetosphere and the pulsar wind zone and, injected
into the pulsar wind nebula should be,
\begin{eqnarray}
\gamma_{Fe}\approx \chi Ze\Phi_{\rm open}/m_{\rm i}c^2\approx
8\times 10^9 \chi B_{12} P_{\rm ms}^{-2},
\label{eq1}
\end{eqnarray}
\noindent
where $m_{\rm i}$ and $Ze$ are the mass and charge of the iron nuclei,
$c$ is the velocity of light, and $\Phi_{\rm open} = \sqrt{L_{\rm rot}/c}$
is the total electric potential drop across the open magnetosphere,
$L_{\rm rot}$ is the rate of rotational energy lost by the pulsar,
$B = 10^{12}B_{12} G$ is the surface
magnetic field of the pulsar, and $P = 10^{-3}P_{\rm ms}$ is the pulsar period.
Due to the rotational energy losses on emission of dipole radiation, the pulsar period
evolves in time according to
\begin{eqnarray}
P_{\rm ms}^2 = P_{\rm 0, ms}^2 + 2\times 10^{-9} t B_{12}^2,
\label{eq2}
\end{eqnarray}
\noindent
where $P_{\rm 0, ms}$ is the  initial period of the pulsar, and $t$ is in seconds.
Arons and collaborators argue that the acceleration
factor $\chi$ is not far from unity. $\chi = 0.5$ is taken in the following
calculations. Moreover, we assume that
these nuclei take significant part, $\xi$, of the pulsar rotational energy.
The unknown value of $\xi$ times the pulsar birth rate in the Galaxy, $\eta$,
can be obtained from the comparison of contribution of particles injected by pulsars
to the observed cosmic ray flux. Eq.~(\ref{eq1}) postulates that the pulsar at
a specific time accelerates nuclei monoenergetically.
However the energies of freshly produced nuclei change in time due to the change of
the pulsar period caused by its rotational energy losses.

We assume that a pulsar can extract iron nuclei from its surface and accelerate
them to high energies during the radio phase. Then
the products of efficient leptonic cascades, occurring in the inner pulsar
magnetosphere, impinge on the polar cap region allowing extraction of
the iron nuclei from the surface.

The iron nuclei extracted from the surface of the pulsar are accelerated in the
outer gaps of the inner pulsar magnetosphere (Cheng et al.~1986), suffering partial
photo-disintegrations in collisions with nonthermal radiation produced in leptonic
cascades. The details of the acceleration model of nuclei in the outer gaps are
given in section 3.1 of Bednarek \& Protheroe~(2002). The level of photo-disintegration
of nuclei depends on the pulsar parameters. To calculate the number of nucleons extracted
from nuclei during propagation through the outer gap we determine the outer gap radiation
field (Bednarek \& Protheroe~2002), by interpolation/extrapolation of the radiation
fields in the Crab type pulsars, calculated by Ho~(1989), and for the Vela type pulsars,
based on the observed change of the $\gamma$-ray luminosity (Ruderman \& Cheng~1988).
Nuclei and protons from their disintegration are additionally accelerated in the pulsar
wind zone to energies postulated by Arons~(1998), see Eq.~1. They
are injected into the pulsar wind nebula. Neutrons from the disintegration of nuclei
move ballistically through the nebula decaying inside or outside it depending on
their Lorentz factor. At the early stage of
expansion, when the nebula is dense enough, nuclei suffer further disintegrations in
collisions with the matter (Bednarek \& Protheroe 2002). These processes are discussed in
the next subsection.

\subsection{Escape of nuclei from expanding nebula}

The nuclei, injected by the pulsar, propagate in the expanding nebula
with parameters changing drastically in time. In order to take properly into account
different effects on their propagation (collisions with matter,
diffusion inside the nebula, escape from the nebula), we have to assume
a time dependent model for the expanding nebula taking into account
not only the initial parameters of the expanding envelope but also
the energy supplied by the pulsar. The model has to determine such basic
parameters as: the expansion velocity of the nebula, the pulsar wind shock radius,
the outer radius of the nebula, the average density of matter and magnetic field strength
inside the nebula, and others.
The details of such a simple model for the PWNe are described in our previous
paper (Bednarek \& Bartosik~2003). We repeat here its main features.

The evolution of a supernova remnant, containing an energetic pulsar, is described
according to the picture considered
by  Ostriker \& Gunn~(1971) and Rees \& Gunn (1974).
Let us denote the initial expansion velocity of the bulk matter in supernova envelope
by $V_{\rm 0,SN}$ and its initial mass by $M_{\rm 0,SN}$.
The  expansion velocity can increase due to the additional supply
of energy to the nebula by the pulsar. It can also decrease due to the
accumulation of the surrounding matter.
We take these processes into account
to determine the radius of the nebula at a specific time, t, by using
the energy conservation,
\begin{eqnarray}
{{M_{\rm SN}(t)V_{\rm SN}^2(t)}\over{2}} = {{M_{\rm 0,SN}V_{\rm 0,SN}^2}\over{2}}
+ L_{\rm pul-neb},
\label{eq3}
\end{eqnarray}
\noindent
where $L_{\rm pul-neb}$ is the part of the energy supplied by the pulsar to the nebula in the form
of magnetic field and the energy transfered from relativistic hadrons to the nebula due to adiabatic
energy losses. The rest of the rotational energy lost by the pulsar is
almost immediately radiated by accelerated leptons. The energy of relativistic hadrons which is
injected into the interstellar medium from the nebula is equal to the difference between the
energy gained by hadrons during acceleration process, assumed equal to $\xi L_{\rm rot}$,
and the energy lost by them to the nebula during its adiabatic expansion.
The part of energy lost by relativistic hadrons during adiabatic expansion of the nebula
has been calculated numerically since the process of diffusion of hadrons inside the nebula
has to be taken into account.

The mass of the expanding nebula increases (by sweeping the surrounding medium) according
to the formula
\begin{eqnarray}
M_{\rm SN}(t) = M_{\rm 0,SN} + {{4}\over{3}}\pi \rho_{\rm sur} R_{\rm Neb}^3(t),
\label{eq5}
\end{eqnarray}
\noindent
where $\rho_{\rm sur}$ is the density of the surrounding medium and $R_{\rm Neb}$
is the outer radius of the expanding envelope at the time, t, which depends on the
expansion history of the nebula,
\begin{eqnarray}
R_{\rm Neb} = \int_0^t V_{\rm SN}(t')dt'.
\label{eq6}
\end{eqnarray}
The expansion velocity of the nebula, $V_{\rm SN}(t)$, and the average density of matter
inside it,
\begin{eqnarray}
\rho_{\rm Neb} = 3M_{\rm SN}(t)/4\pi R_{\rm Neb}^3(t),
\label{eq6a}
\end{eqnarray}
\noindent
at an arbitrary time have been found by solving numerically the above set of Eqs. (3-6).
Note however, that the average density of matter, which is very useful for our simple
model,  may not correspond to the real
distribution of matter inside the nebula specially at the later phase of expansion when the
part of matter can form dense filaments. For example the matter inside the Crab Nebula
is accumulated in filaments containing $4.6\pm 1.8$ M$_{\odot}$ of ionized and
neutral gas (Fesen et al.~1997) with density $\sim 500$ cm$^{-3}$ (Davidson \& Fesen~1985).
No significant gas has been confirmed around the Crab Nebula (Fesen et al.~1997).
Atoyan \& Aharonian~(1996) suggest that relativistic hadrons
can be efficiently trapped by these dense filaments.
If this is the case then the efficiency of interactions of relativistic nuclei with the matter
can be enhanced in respect to the interactions with the uniformly distributed matter inside
the nebula.

\begin{figure*}
  \vspace{11.5truecm}
\includegraphics{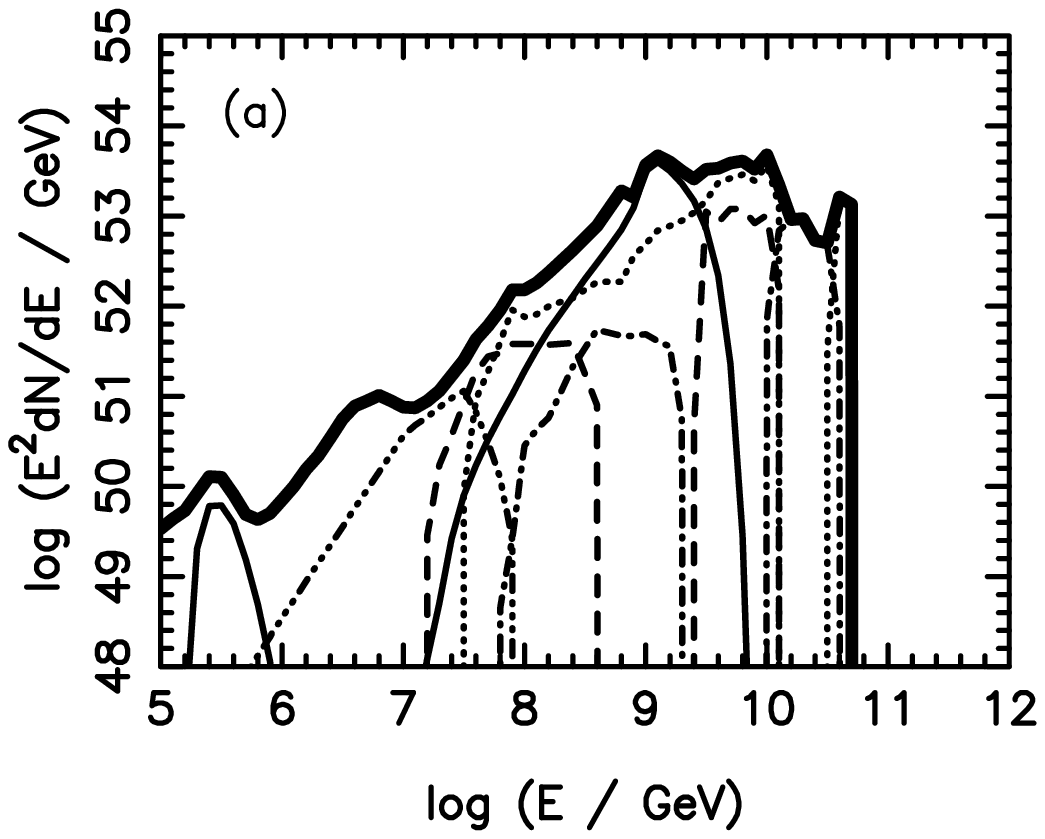}
\includegraphics{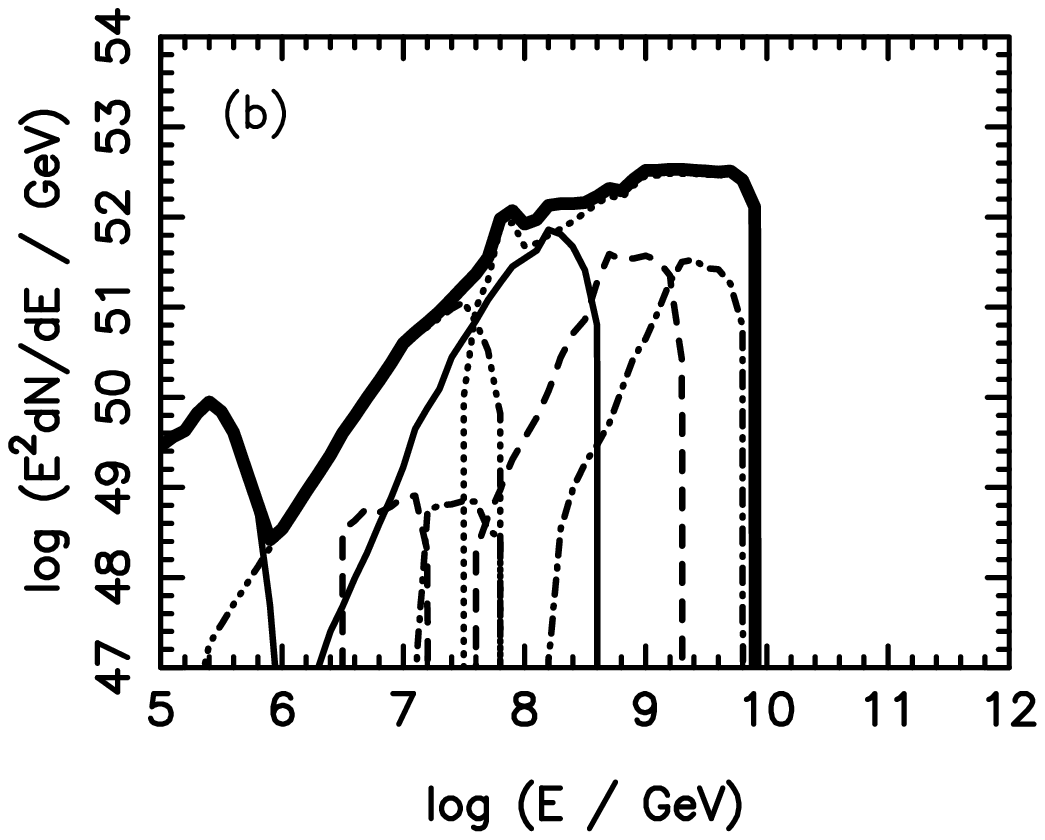}
\includegraphics{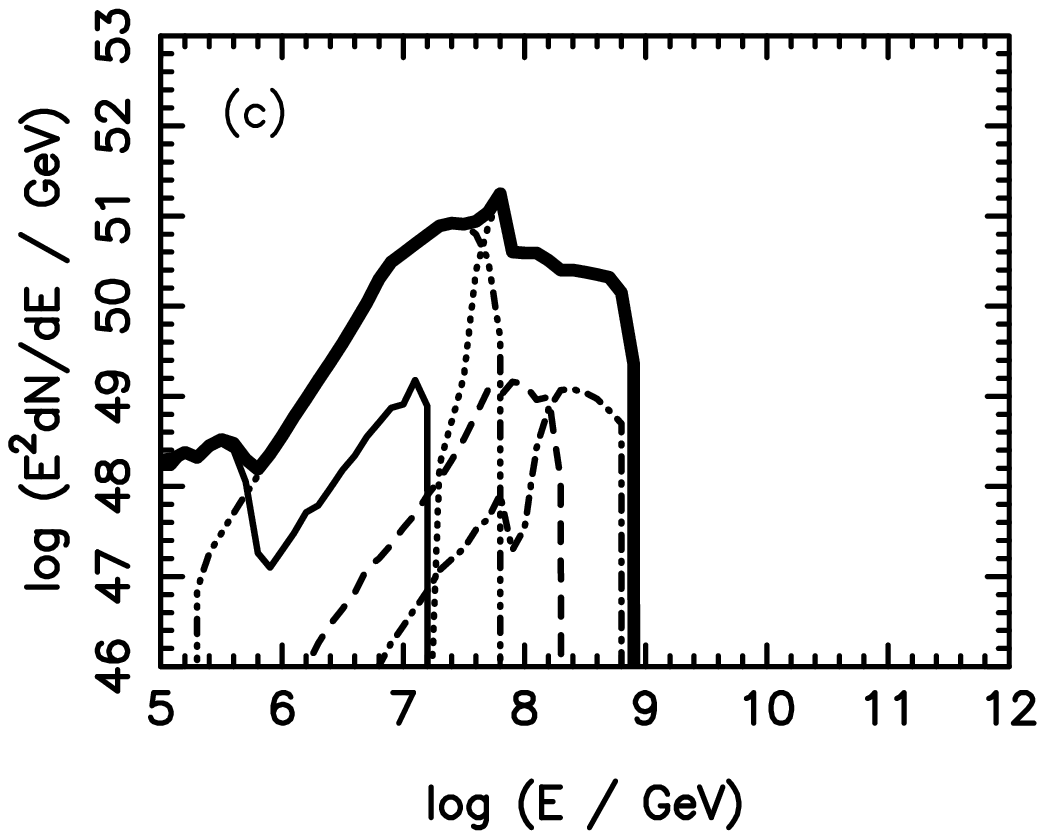}
\includegraphics{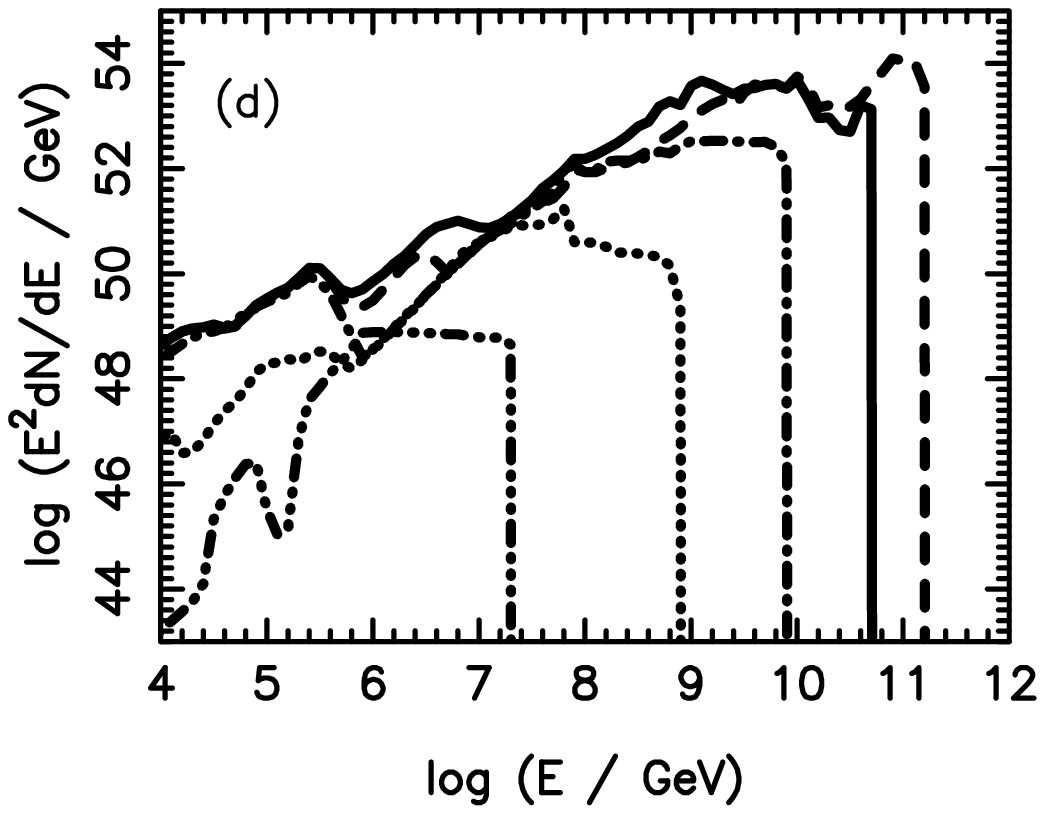}
  \caption{Energy spectra of different types of nuclei escaping from a supernova
nebula into the galactic medium. The nuclei are accelerated by the pulsar with
the surface magnetic field $B = 4\times 10^{12}$ G and different initial
periods (a) 1 ms, (b) 10 ms, and (c) 100 ms. The pulsar has been born during the
supernova explosion with the envelope mass of $4 M_{\odot}$ expanding with the
initial velocity 2000 km s$^{-1}$. The total spectrum of nuclei is marked by the thick full
curve. Specific curves show the spectra of nuclei with
the mass numbers A = 56 (dot-dot-dot-dashed), 41-55 (dotted), 11-40
(dot-dashed), 2-10 (dashed), protons
and  protons from decay of neutrons (thin full). The spectra of nuclei
for pulsars with different initial periods
1 ms (full curve), 10 ms (dot-dashed), $10^2$ ms (dotted), and $10^3$ ms
(dot-dot-dot-dashed) are shown in (d).
The case of the pulsar with 1 ms initial period and the expansion velocity of
$10^4$ km s$^{-1}$ is shown by the dashed curve.}
\label{fig1}
\end{figure*}

The pulsar loses energy in the form of a relativistic wind extending up to the
shock at a distance $R_{\rm sh}$. At this distance, the pressure of the wind
is balanced by the pressure of the expanding nebula. Rees \& Gunn~(1974) estimate
the location of this
shock as a function of time by comparing the pulsar wind energy flux, determined by
$L_{\rm rot}$ (see Eq.~\ref{eq4}), with the
pressure of the outer nebula, determined by the supply of energy to the
nebula by the pulsar over the whole of its lifetime,
\begin{eqnarray}
{{L_{\rm rot}(t)}\over{4\pi R_{\rm sh}^2c}}\approx
{{L_{\rm pul-neb}}\over{{{4}\over{3}}\pi R_{\rm Neb}^3}}.
\label{eq7}
\end{eqnarray}
\noindent
and
\begin{eqnarray}
L_{\rm rot}(t) = B_{\rm s}^2 R_{\rm s}^6 \Omega^4/6c^3\approx
2.5\times 10^{43}B_{12}^2P_{\rm ms}^{-4}~~{\rm erg~s}^{-1},
\label{eq4}
\end{eqnarray}
\noindent
where $R_{\rm s}$ and $B_{\rm s}$ are the radius of the pulsar and its surface magnetic
field, $\Omega = 2\pi/P$. The pulsar period changes according to Eq.~\ref{eq2}.
It is assumed in this formula that $R_{\rm sh} \ll R_{\rm Neb}$.

Knowing how the magnetic field depends on the distance from the pulsar in the pulsar
wind zone, we can estimate the strength of the magnetic field at the shock region
from
\begin{eqnarray}
B_{\rm sh} = \sqrt{\sigma} B_{\rm s}\left({{R_{\rm s}}\over{R_{\rm lc}}}\right)^3
{{R_{\rm lc}}\over{R_{\rm sh}}},
\label{eq8}
\end{eqnarray}
\noindent
where $\sigma$ is the ratio of the magnetic energy flux to the particle energy
flux from the pulsar at the location of the pulsar wind shock. There is
observational evidence that $\sigma$ evolves in time. It is much less than one for
very young nebulae (of the Crab type), and increases with age being of the order of
one for nebulae of the Vela type. The value of $\sigma$ for a specific pulsar is
determined by the efficiency of conversion of magnetic energy into the particles energy
in the pulsar wind zone. The evolution of
$\sigma$ with the parameters of the pulsar is
estimated by interpolating between the values for the Crab pulsar,
equal to $\sim 0.003$ (Kennel \& Coroniti~1984), and for the Vela pulsar, equal to
$\sim 1$ (Helfand et al.~2001).
We assume that the magnetic field strength in the volume of the nebula
depends on the distance from its center as described by the formulae
derived in the magnetohydrodynamic model for
the PWNe considered by Kennel \& Coroniti~(1984).
The geometry of the magnetic field inside the nebula may have strong effect on the
diffusion of nuclei inside it. Recent 3-dimensional modelling of the Chandra observations of the
Crab Nebula suggests that the regular component of the magnetic field should be comparable to the
turbulent component (Shibata et al.~2003). In our calculations the diffusion in turbulent component
is only considered. The presence of regular toroidal magnetic field should slow down the diffusion
of nuclei through the nebula.

It is assumed that nuclei injected into the nebula at a specific time $t_{\rm inj}$,
escape from it at the time $t_{\rm esc}$, if their diffusion distance in the turbulent magnetic
field of the nebula, $R_{\rm diff}$, is equal to the dimension of the nebula,
$R_{\rm Neb}$, at the time $t_{\rm esc}$. The diffusion distance in the magnetic field
of the nebula is obtained by integration (Bednarek \& Protheroe~2002),
\begin{eqnarray}
R_{\rm diff} = \int^{t_{\rm esc}}_{t_{\rm inj}} \sqrt{{{3D}\over{2t'}}}dt',
\end{eqnarray}
\noindent
where the diffusion coefficient is taken to be $D = R_{\rm L}c/3$, and
$R_{\rm L}$ is the Larmor radius of nuclei depending on the distance from the center of the
nebula. $R_{\rm diff}$ and $t_{\rm esc}$ have been calculated numerically.
We include also the adiabatic losses
of nuclei during the period of their propagation inside the nebula.
Due to this losses and interaction of nuclei with the matter the energy
of a nucleus at a specific time is

\begin{eqnarray}
E(t) = E(t_{\rm inj}) {{t_{\rm inj} + t}\over{2t_{\rm inj}K^\tau}}
\end{eqnarray}
\noindent
where $E(t_{\rm inj})$ is its energy at
the time $t_{\rm inj}$, K is the inelasticity coefficient for collisions
of nuclei with the matter,
and $\tau$ is the optical depth calculated based on the known density of
matter inside the nebula (see Eq.~7).
The details of these calculations are given in sections 4.2 and 4.3 in Bednarek \&
Protheroe~(2002).

\begin{figure*}
  \vspace{11.5truecm}
\includegraphics{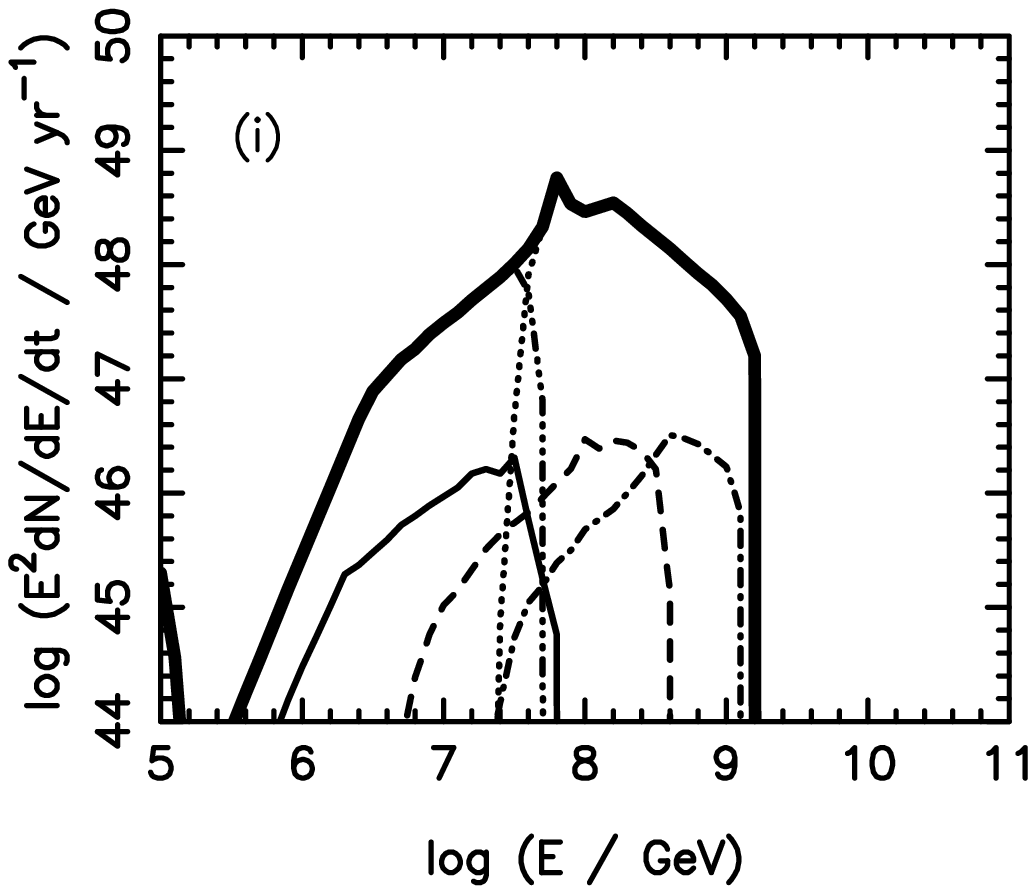}
\includegraphics{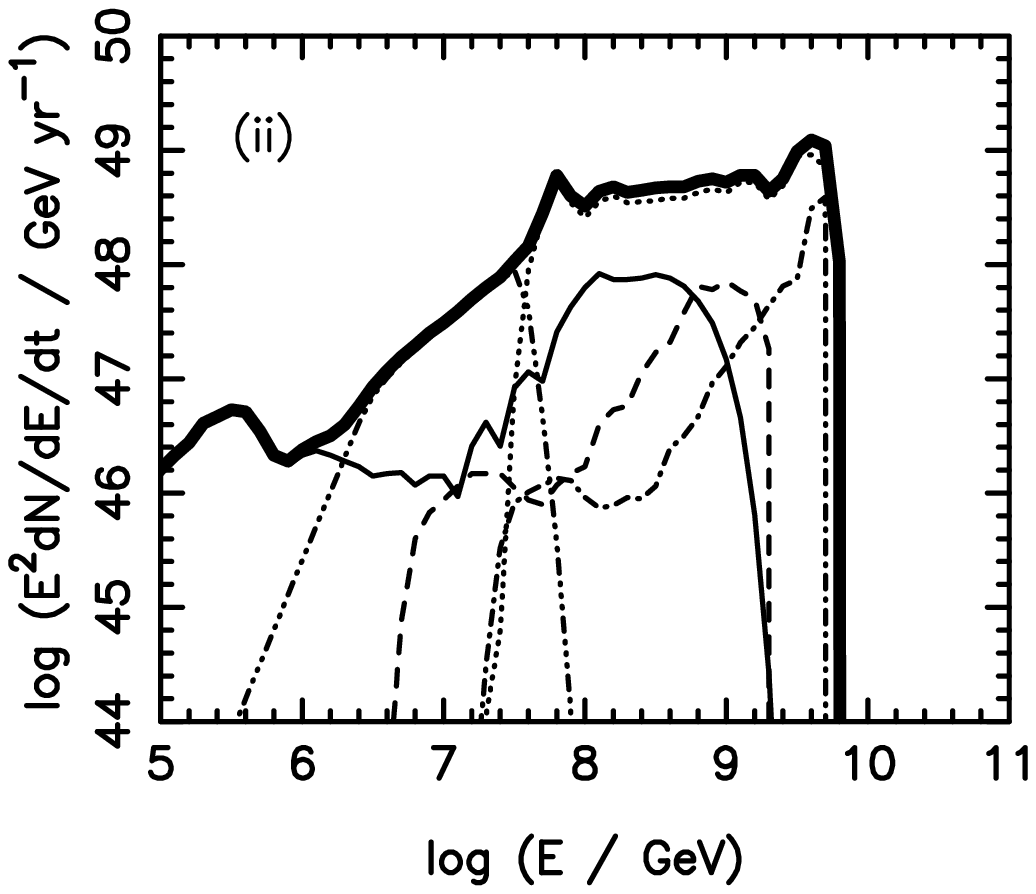}
\includegraphics{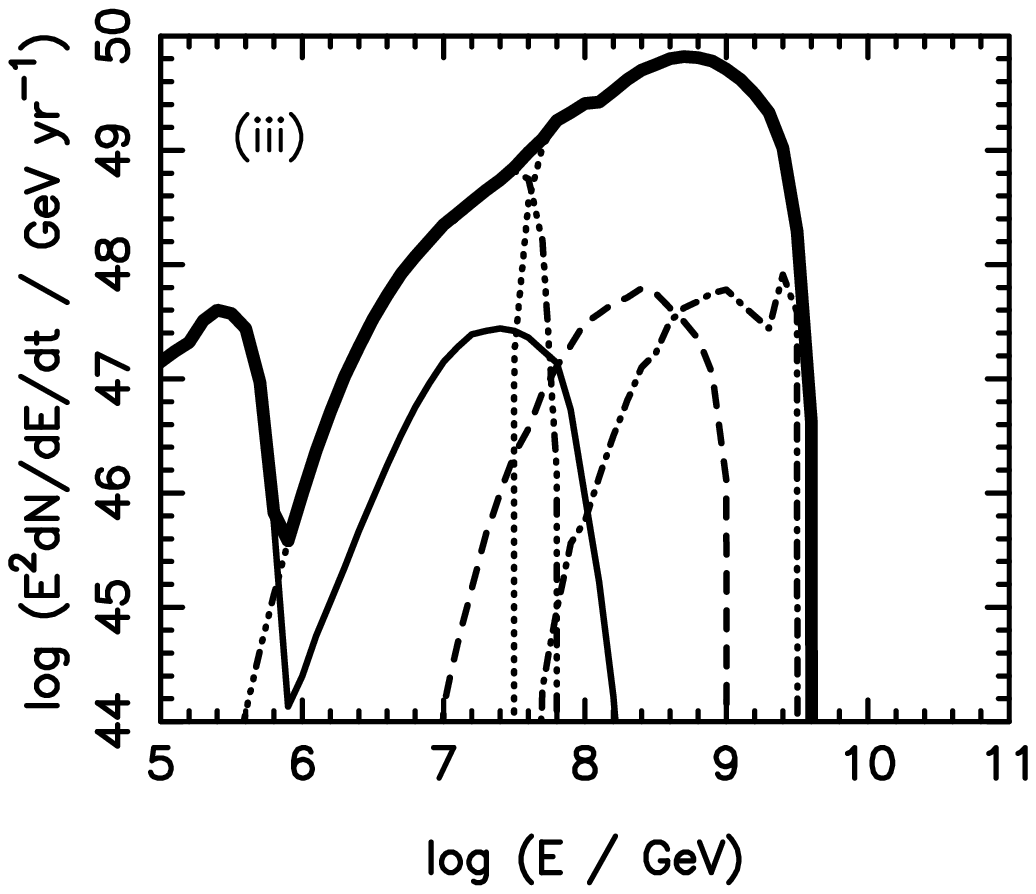}
\includegraphics{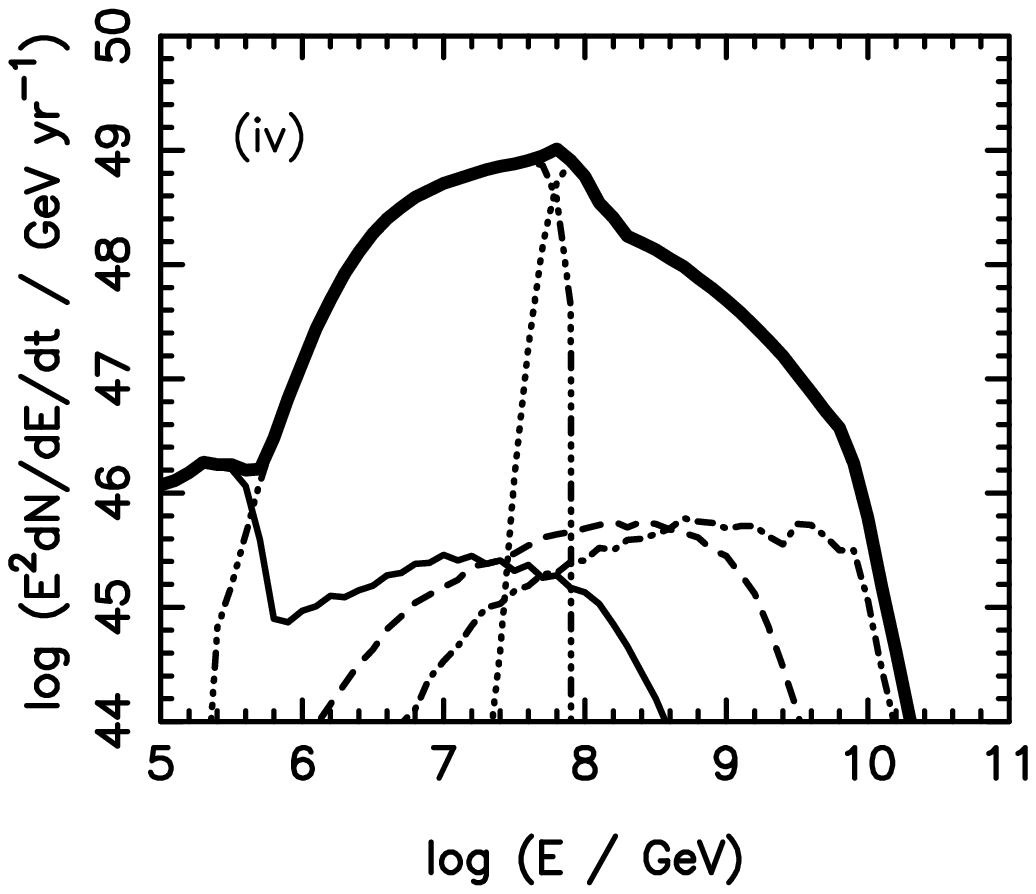}
  \caption{{The spectra of nuclei, injected by the galactic population of pulsars
into the Galaxy, calculated for different models
of initial parameters of the pulsar population, obtained in the work by
Narayan~(1987) and van der Swaluw \& Wu~(2001) (model i), Narayan~(1987) and
Xu et al. (2001) (model ii), Model A of Lorimer et al.~(1993) (our model iii), and
Model B of Lorimer et al.~(1993) (model iv) (see text for details). The spectra
of different types of nuclei are marked as in Figs.~1.}}
\label{fig2}
\end{figure*}
Considering all the
processes discussed above, we calculate the energy spectra of different groups of nuclei
escaping from the nebula into the Galaxy for arbitrary initial parameters of the
pulsars. Protons from decay of neutrons, released
from heavy nuclei during their propagation through the radiation field of the outer gap
of the pulsar and later as a result of collisions with the matter inside the nebula,
are also included in
these calculations. Neutrons extracted from nuclei in collisions with the
outer gap radiation field have relatively low Lorentz factors since their parent nuclei
are pre-accelerated in the
outer gap to energies lower than that achieved later in the pulsar wind zone.
On the other hand, neutrons extracted from nuclei in collisions with the matter
have typically higher Lorentz factors than protons since, moving balistically through the
PWNa, do not lose energy on the adiabatic expansion. We assume that neutrons decaying
within 3 kpc from the galactic disk are captured by the magnetic field of the galactic
halo and contribute to the galactic CRs.
As an example, Figs.~1 show the spectra of different types of nuclei injected into the
Galaxy for the case of a supernova with initial parameters derived for
the Crab Nebula, i.e. the mass of the supernova envelope equal to 4 M$_{\odot}$,
the initial expansion velocity 2000 km s$^{-1}$, and
the density of surrounding medium equal to 0.3 cm$^{-3}$.
Figs.~1a-c show spectra of the escaping
nuclei within selected range of mass numbers for pulsars with initial periods
1 ms, 10 ms, and 100 ms, and a typical surface magnetic field of
$4\times 10^{12}$ G. It is clear
that nuclei injected by pulsars with shorter initial periods suffer more severe
fragmentations. The total spectra of these secondary nuclei can be even
dominated in some range of energies by secondary protons extracted from heavy
nuclei and from decay of neutrons (thin full curves).
The total spectra of the nuclei above $\sim 10^{(7-8)}$ GeV have the spectral index not
far from -2 extending through about two decades in energy. The low energy bump in the
spectra below $\sim 10^6$ GeV is due to the neutrons extracted from nuclei during their
fragmentation in the radiation field of the outer gap of the pulsar.
Fig.~1d compares the total spectra of nuclei injected by pulsars with
different initial periods and initial expansion velocities of the
supernova envelopes. Pulsars with longer initial periods should be able to contribute
to lower energy range in the total spectrum of nuclei. On the other hand,
larger initial expansion velocity of the envelope allows the nuclei to escape
at earlier times because the magnetic field inside the envelope drops more rapidly.
Therefore, these nuclei escape with larger energies due to their smaller energy
losses on the adiabatic expansion.

\subsection{Pulsar population within the Galaxy}

The spectrum of nuclei injected inside the Galaxy can be obtained by summing up over
all population of pulsars inside the Galaxy. However, the initial parameters of
the pulsars are not precisely known. Therefore, we consider a few 
different models proposed in the literature. They differ in distributions of 
the surface magnetic fields and the initial periods of the new born pulsars.
The following models are:

\begin{enumerate}

\item The surface magnetic fields of the pulsars are described by the distribution
derived by Narayan (1987),
\begin{eqnarray}
dN/d(log B)\approx 0.065 / (B [10^{12} G])~{\rm year^{-1}},
\end{eqnarray}
\noindent
for $B > 2\times 10^{12}$ G. The distribution of pulsars with the surface magnetic
fields below $2\times 10^{12}$ G bases on Fig.~13 in Narayan~(1987).
All pulsars are born with the fixed initial period $P_{\rm 0} = 40$ ms
(van der Swaluw \& Wu~2001).

\item  The surface magnetic fields of the pulsars as in model (i).
The initial pulsar periods are correlated with their surface magnetic fields,
\begin{eqnarray}
P_{\rm 0} [ms] = 63.7/(B [10^{12} G]),
\end{eqnarray}
\noindent
as postulated by Xu et al.~(2001). We apply this formula for the pulsar initial periods
above 2 ms.

\item  Model A of Lorimer et al.~(1993) postulating the Gaussian distribution of
log~B of the pulsars with  parameters $<log B [G]> = 12.46$ and $\sigma_{\rm log B} =
0.31$. We apply this model for the range of $B=10^{11.5-13.3}$ G.
All pulsars are born with a single  initial period, $log P_{\rm 0} [ms] = 1.35$.

\item Model B of Lorimer et al.~(1993) postulating the Gaussian distributions of
log~B, with $<log B [G]> = 12.3$ and $\sigma_{\rm log B} = 0.35$, and for
log $P_{\rm 0}$, with $<log P_{\rm 0} [ms]> = 2.6$ and $\sigma_{\rm log B} = 0.35$
above 2 ms.

\end{enumerate}

\begin{figure}
  \vspace{11.5truecm}
\includegraphics{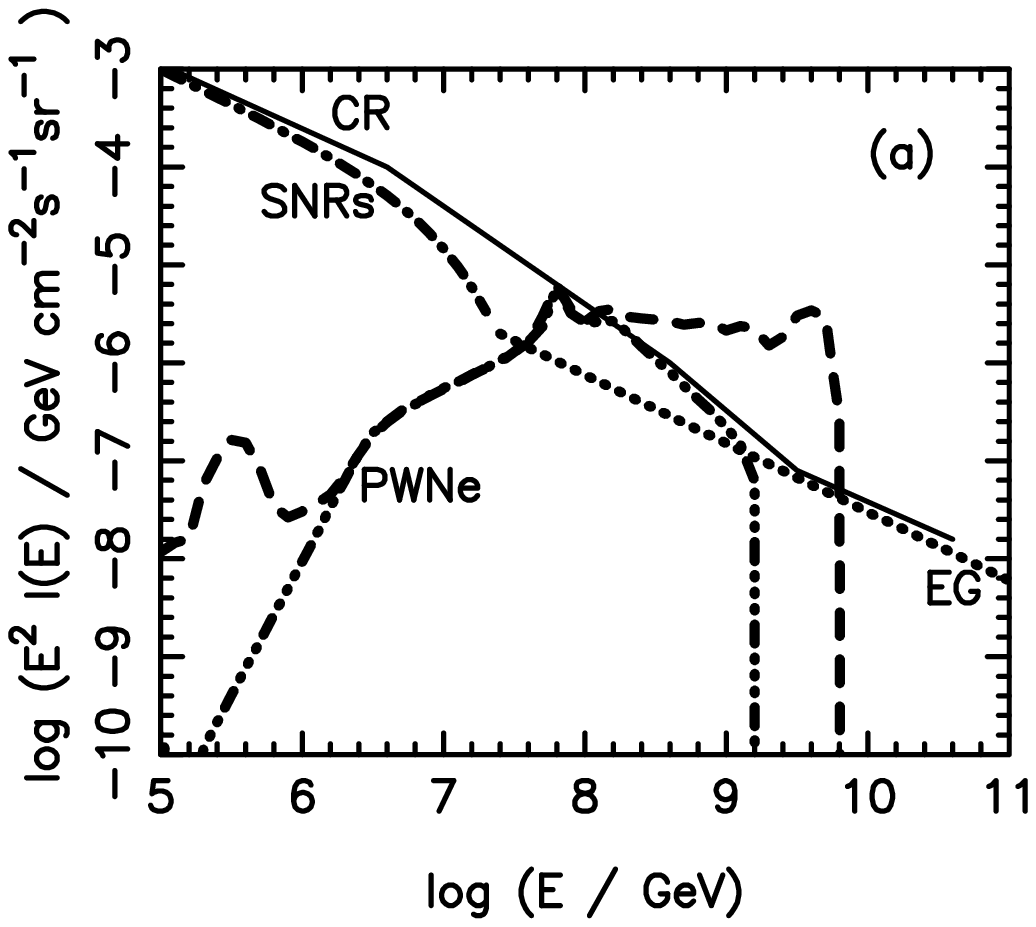}
\includegraphics{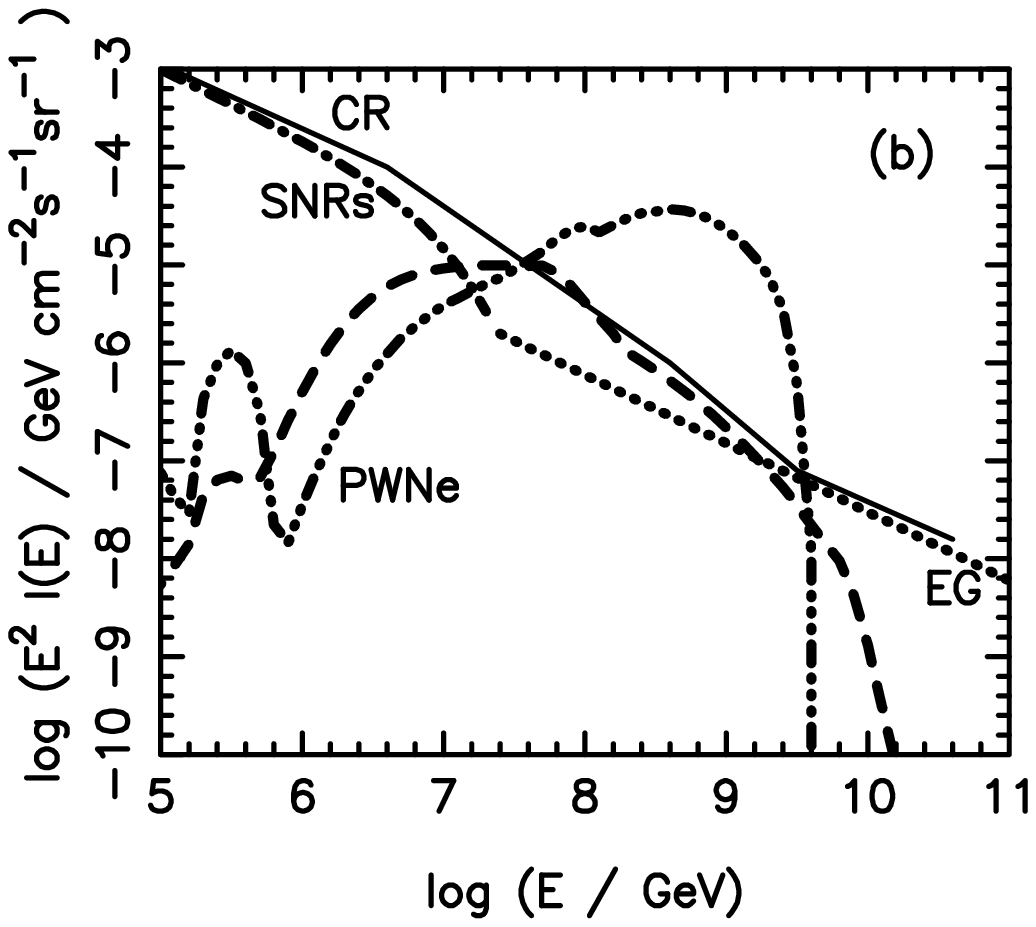}
  \caption{The comparison of the spectrum of particles accelerated by pulsars
with the parameters described by the model (i) (dot-dot-dot-dashed curve in (a)),
model (ii) (dashed curve in (a)), model (iii) (dot-dot-dot-dashed curve in (b)),
and model (iv) (dashed curve in (b)), with the observed cosmic ray spectrum
(thin full curve). It is assumed that at low energies the supernova remnants (SNR)
accelerate CRs with the
spectrum $dN/dE\propto E^{-2.7}exp(-E/10^7 GeV)$ (dot-dashed curve). At
extremely high energies (EG), the spectrum has the form $dN/dE\propto E^{-2.7}$
(dotted line).}
\label{fig3}
\end{figure}

For these four models of the pulsar population, we calculate the injection spectra of
nuclei into the Galaxy applying the values of the normalization coefficients,
$\xi\cdot \eta$, which give us the flux of calculated nuclei on the level comparable to
that observed in the cosmic ray spectrum.  
It is assumed that the mass of nebula surrounding the pulsar is equal to 3 M$_\odot$ and
expends with the velocity of 2000 km s$^{-1}$. 
Model (i), calculated for $\xi\cdot\eta = 1/700$ yr$^{-1}$,
gives a power law spectrum of nuclei over about two decades above $\sim 10^8$ GeV
and very sharp cut-off at lower energies (see Fig.~2i). The spectrum is dominated by
heavy nuclei from the iron group.
Model (ii), calculated also for $\xi\cdot\eta = 1/700$ yr$^{-1}$, gives very flat
spectrum at high energies with the spectral index close to 2 (see Fig.~2ii).
The primary iron nuclei disintegrate significantly in this model.
Protons from decay of neutrons can contribute to the total spectrum at around 
$10^{18}$ eV. These features are due to a relatively large number of pulsars with short
periods predicted by this model.
Model (iii), calculated for $\xi\cdot\eta = 1/120$ yr$^{-1}$, gives the total
spectrum with a rather narrow, strong peak at $\sim 10^8-10^9$ GeV (see Fig.~2iii).
The total spectrum is dominated by heavy nuclei from the iron group.  Such shape of the
spectrum is due to the assumption about the fixed period of all new born pulsars
which is probably not very realistic.
Model (iv), $\xi\cdot\eta = 1/120$ yr$^{-1}$, predicts the power law spectrum of
nuclei between a few $10^7$ GeV and  $10^{10}$ GeV with the index close to 3
(see Fig.~2iv). The spectrum is dominated by heavy nuclei in all this energy range.
In the next section we compare the spectra of nuclei calculated in terms of these
models with the observed spectrum and the mass composition of CRs.

\section{Contribution to the cosmic rays in the Galaxy}

Having in hand the injection spectrum of nuclei by pulsars, we can estimate
the flux of these nuclei inside the Galaxy adopting the leaky box model
from the relation

\begin{eqnarray}
{{dN}\over{dE dS dt d\Omega}} = {{dN_{\rm inj}}\over{dE dt}}\cdot
{{c~\tau_{\rm esc}}\over{4\pi ~V_{\rm gal}}},
\label{eq9}
\end{eqnarray}

\noindent
where $dN_{\rm inj}/dEdt$ is the total injection spectrum of CRs from all the pulsars
in the Galaxy (calculated in Sect.~2),
$V_{\rm gal} = 10^{68}$ cm$^3$ is the volume of the Galaxy (a disk with a
radius of 15 kpc and a half-thickness of 3 kpc), and $\tau_{\rm esc}$
is the escape time of hadrons from the Galaxy. The dependence of the escape time of
nuclei on their energy and charge is approximated by

\begin{eqnarray}
\tau_{\rm esc} = {{2\times 10^{7}~yr}\over{(E_{\rm GeV}/Z)^{0.3}}},
\label{eq10}
\end{eqnarray}

\noindent
where $Z$ is the charge of nuclei and $E_{\rm GeV}$ its energy in GeV.
This formula is normalized to
the estimated lifetime of protons with energies $10^{18}$ eV equal to $4\times 10^4$
years (see Fig.~4.18 in Berezinsky et al.~1990).

The calculations have been performed for all four models of the pulsar population,
applying the normalization parameter, $\xi\cdot\eta$, as reported above.
Our aim is to find out if any of the proposed so different pulsar population 
models is able to describe general features of the observed cosmic ray spectrum.
The comparison with the CR and calculated spectra are shown
in Figs.~3. When fitting the observed CR spectrum in the broad energy range,
we assume the popular hypothesis suggesting that the shape of the observed
cosmic ray spectrum is a combination of four different components. Below the knee 
the power law spectrum has the form $\propto E^{-2.7}$. It is widely believed that
the main contribution to this part of the spectrum comes from the particles accelerated in the supernova
shock waves resulting in a power law spectrum and the exponential cut-off which we put at
$10^7$ GeV, $dN/dE\propto E^{-2.7}exp(-E/10^7 GeV)$. Between the knee and the second 
knee at a few $10^{17}$ eV, the spectrum  has the form $E^{-3}$,
and between the second knee and the ankle $\propto E^{-3.2}$.
The extremely high energy part of the cosmic ray spectrum is
due to another component approximated by a simple power law with the
spectral index -2.7. Particles, accelerated by the pulsars, dominate the spectrum
between the knee and the ankle. 

However, only model (iv) gives good general consistency of the observed shape 
of the spectrum with the calculated one.  Model (i) predicts
significant deficit of particles at $\sim 10^{16}$ eV, which has not been reported by
experimentalists. This deficit might be suppressed if the pulsars are able to accelerate
nuclei for longer time than the radio activity period expected from the modeling of
magnetospheric high energy cascade processes.
Models (ii) and (iii) predict too flat spectra between the knee and the ankle and
thus should be rejected.

We also calculate the expected mass composition of the CRs, $<ln A>$, in terms of
the model (iv) and compare it with the measurements of the mass composition
of CRs by different experiments (see Fig.~4).
The mass composition of the galactic supernova component is taken as reported
by the direct measurements at low energies (JACEE, RUNJOB). The same composition is
also applied to the extremely high energy component. 
We calculate the average mean $<lnA>$ for our model (iv) 
assuming two models for extend of the extragalactic (EG) component. 
In the first one, EG component extends down to at least $3\times 10^7$ GeV. 
In the second one, it breaks at $10^9$ GeV (see Fig.~4).  
The Model (iv) with EG component extending to low energies describes well
the general tendency reported by most experiments, i.e. the rise of
$<ln A>$ above the knee, and gradual decrease of $<ln A>$ above 
$\sim 3\times 10^{17}$ eV. Therefore, we conclude that
the model of the pulsar contribution to the cosmic ray spectrum
can explain the change of the average mass above the knee, which is
the consequence of a sudden switching off the pulsar acceleration mechanism of heavy
nuclei due to the lack of their extraction from the neutron star surface.
In our model this is due to the switching off the process of electromagnetic cascading
in the inner pulsar magnetosphere whose products heat up the polar regions of
the neutron star. If the light extragalactic component does not extend to
low energies but breaks at $\sim 10^{18}$ eV, then our model 
predicts rather heavy composition up to $\sim 10^{18}$ eV as suggested by some groups 
(Dawson et al.1998, Ave et al.~2003, Dova et al.~2003).

\begin{figure}
  \vspace{6.5truecm}
\includegraphics{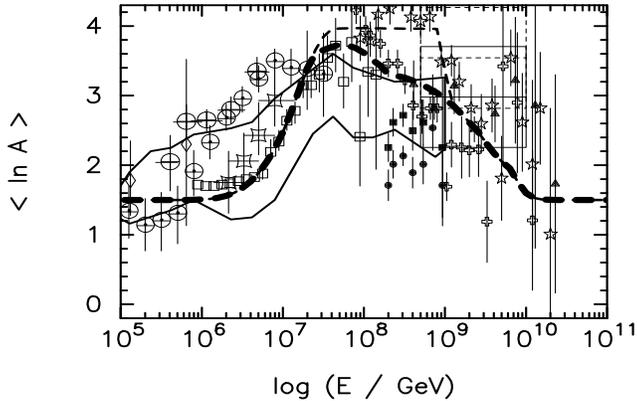}
  \caption{The comparison of the mass composition of the cosmic rays, $<ln A>$,
obtained in terms of the model (iv) for the case of the light extragalactic
component extending to low energies (thick dashed curve) and breaking at $10^{18}$ eV
(thin dashed curve), with the measurements reported by different experiments
(JACEE - Shibata et al.~(1999), RUNJOB - Apanasenko et al.~(2001),
CASA-MIA - Glasmacher et al.~(1999), KASCADE (e/m) - Ulrich et al.~(2001),
KASCADE (p/m) - Engler et al.~(1999), EAS-TOP - Alessandro et al.~(2001),
FLY'S-EYE - Dawson et al.~(1998),
Akeno A1 - Hayashida et al.~(1995), Akeno A100 - Hayashida et al.~(1997), Haverah Park
- Ave et al.~(2003), Volcano Ranch - Dova et al.~(2003)). The region between the thin 
full curves shows $<ln A>$ obtained from many experiments and derived by 
H\"orandel~(2004). $<ln A>$ equal to 1.5 is taken for the SNR contribution to cosmic ray
spectrum as measured at low energies by direct experiments. Similar composition is also
applied for the extragalactic (EG) component of CRs at the highest energies.}
\label{fig4}
\end{figure}
\section{Discussion and Conclusions}

The best fit to the observed cosmic ray spectrum between the knee and the ankle
is obtained for the model B of Lorimer et al.~(1993) (our model iv) which postulates
that the observed radio pulsars are born with relatively long initial periods, with the
average value of $\sim 400$ ms, with the Gaussian distribution in log scale, and
typical surface magnetic fields, with the average value of $2\times 10^{12}$ G and also
the Gaussian distribution in log scale. This average value of the pulsar periods at 
birth is similar to that one obtained in the analytical work by Giller \& Lipski~(2002), 
who got the best description of the cosmic ray spectrum for the gamma
distribution of the initial pulsar periods (which is $\propto P_{\rm 0}^{s-1}$
for small periods) with average value of $P_{0} = 500$ ms and s = 3.86,
and the Gaussian distribution for log B with average value between $10^{12}-10^{13}$ G
and $\sigma_{\rm log B} = 0.4$. However, the shape of the pulsar distribution for initial 
periods is very different from that one derived by Lorimer et al.~(1993), and applied in 
our work. This is not surprising since, in contrast to Giller \& Lipski~(2002), we
take into account the propagation and escape conditions, and adiabatic and collisional
energy losses of nuclei during their passage throughout the pulsar wind nebula. 
These processes have strong effects on the injection spectrum of nuclei into the Galaxy.
From the normalization of the calculated spectrum of nuclei in terms of the model (iv)
to the observed cosmic ray spectrum, we obtained the efficiency of conversion of the
rotational energy of the pulsar into the energy
of iron nuclei multiplied by the pulsar birth rate equal to $\sim 1/120$ yr$^{-1}$.
This value is consistent with the estimated pulsar birth rate $\sim 1/(20 - 250~{\rm yr})$
(Lyne et al.~1985; Lorimer~2003, Vranesevic et al.~2003). Therefore, we conclude that
the efficiency of acceleration of the iron nuclei should be in the range $\sim
(0.2 - 1.0)$.

The low energy break in the spectrum of nuclei injected by the PWNe is explained in our
model
by switching off the mechanism of extraction of the iron nuclei from the surface of the
neutron star. This is caused by the lack of efficient cascading in the inner pulsar
magnetosphere which products impinge on the region of the polar
cap and allow efficient extraction of the iron nuclei from the neutron star surface.
The cascades stop developing in the pulsar magnetosphere when the pulsar period
becomes too low due to the pulsar rotational energy losses.
This happens when pulsars reach
the age of $\sim 10^6-10^7$ yrs, depending on the value of their surface magnetic field.
Note however, that Giller \& Lipski~(2002) explain this low energy cut-off
by the decay of the surface magnetic field of the pulsars on a typical time scale of
$\sim 10^7$ yrs.

The models (i) and (iii) for the initial parameters of the pulsar
population are not consistent with the observed spectral shape of the CRs, since they
give too narrow spectra with the peak somewhere
between the knee and the ankle. Model (ii), postulating a correlation between
the initial pulsar period and the strength of the surface magnetic field of the the
pulsar (Xu et al. 2001), overproduce CRs in the region of the ankle
by about an order of magnitude, even for the assumed here relatively strong dependence of
the lifetime of particles inside the Galaxy on their energy.

The numerical calculations presented here allow us to include the energy loss processes
of nuclei,
accelerated by the pulsar, during their propagation inside the PWNe and also consider
their escape conditions from the PWNe. Therefore, we can predict the
mass composition of nuclei injected from the PWNe into the Galaxy and
compare it with the observations. The mass composition estimated in terms of our best
fit model (iv) shows rise in the energy range $\sim 10^6-10^7$ GeV, and slower fall 
between $\sim 10^8$ GeV and a few $10^9$ GeV, provided that the light extragalactic 
component expected at the highest energies extends down to $3\times 10^7$ GeV.
This behaviour of the mass composition is in general agreement with the most 
observational results available in the 
literature and collected in the paper by H\"orandel (2004) (see Fig.~4).
However, if the extragalactic component breaks at $10^9$ GeV, then our model 
predicts heavy composition below $10^9$ GeV as suggested by some other experiments
(Dawson et al.1998, Ave et al.~2003, Dova et al.~2003).

The models considered by us have also other interesting, worth to be mentioned features.
For example, model (ii), postulating that a significant part of pulsars are
born with short initial periods and strong surface magnetic fields (correlation between
these two parameters), predicts domination of the mass composition
at energy $\sim 10^9$ GeV by protons from decay of neutrons released from nuclei inside
the PWNa. Note that the existence of an additional pulsar population inside the Galaxy
with extreme parameters seems to be required by the
observations of anomalous X-ray pulsars and magnetars. It can be responsible for the
observed anisotropies of CRs from directions of the Galactic Centre and
the Cygnus region (see Bednarek~2002, Bednarek~2003). This pulsar population might be
also responsible for the highest energy component of the CR
spectrum ($>10^{19}$ eV), provided that heavy nuclei are able to escape from the PWNe
(see Blasi et al.~2000, and Fig.~1d in this paper). Such model predicts very
heavy composition of the highest energy CRs.

It might be surprising that we postulate a similar order contribution to the cosmic ray
spectrum around the knee region from particles accelerated at the outer supernova shock
waves and from the pulsars. However, it is likely that the
initial parameters of pulsars and expanding supernova envelopes, which originate in this
same phenomenon, are in some way related.
Pulsars with more extreme parameters (initial periods, surface magnetic fields) seem to
be produced by explosions of type Ib/c supernovae which progenitors rotate fast and
have relatively smaller envelopes just before explosion.
Therefore, their envelopes have larger velocities, expanding to larger
distances and producing outer shocks with larger dimensions than in the case of more massive
type II supernovae. Such shocks allow more efficient particle acceleration.
Moreover, the exact shape of the cosmic ray spectrum in the knee region is not precisely
known leaving place for some speculations concerning its smoothness.
For example, Erlykin \& Wolfendale (1997) argue for the existence of fine
structures in the knee region of the cosmic ray spectrum to be interpreted as
contribution from nuclei with different mass numbers.

The rate of pulsar formation inside the Galaxy, estimated on 
$\sim 1/(20 - 250~{\rm yr})$ (Lyne et al.~1985; Lorimer~2003, Vranesevic et al.~2003),
is about three orders of magnitude higher than the GRB event rate in our Galaxy. 
These pulsars can supply hadrons with the rate required by the observed density of 
the CRs in the Galaxy. Therefore, the population of classical radio pulsars observed 
in the Galaxy can provide at least similar contribution to the galactic CRs as recent 
GRB model considered by Wick et al.~(2004).

\begin{acknowledgements}
We would like to thank M. Giller and the anonymous referee for 
many useful comments. This work is supported by the Polish KBN grants 
No. 5P03D 025 21 and PBZ-KBN-054/P03/2001.
\end{acknowledgements}

\end{document}